# Performance Degradation Assessment for Electrical Machines Based on SOM and Hybrid DHMM

**Chong Bian[1], Shunkun Yang*, [2], Tingting Huang[2], Qingyang Xu[3], Jie Liu[2], Enrico Zio[4, 5, 6]**
1.School of Automation Science and Electrical Engineering, Beijing University of Aeronautics and Astronautics, Beijing 100191, China
2.School of Reliability and Systems Engineering, Beijing University of Aeronautics and Astronautics, Beijing 100191, China
3.Infrastructure Inspection Research Institute, China Academy of Railway Sciences, Beijing 100191, China
4.Chair on system science and energetic challenges, EDF Foundation, Laboratoire Genie Industriel, CentraleSupélec, Université Paris-Saclay, 3 rue Joliot Curie, 91190 Gif-sur-Yvette, France
5.Energy Department, Politecnico di Milano, Piazza Leonardo da Vinci, 32, 20133 Milano, Italy
6.Sino-French Risk Science and Engineering Lab, Beijing University of Aeronautics and Astronautics, Beijing 100191, China

*Abstract*—Aimed at the timely detection of the degradation of electrical machines and the organization of active maintenance, numerous studies on performance degradation assessment have been conducted. However, previous research still suffers from two deficiencies: 1) determining the relevant relationship among diverse machine degradation states and assessing the specific degree of deterioration and 2) determining the evolutionary relationships among degradation and failure modes and assessing the failure modes corresponding to different degradation scenarios. To address these two deficiencies, a novel performance degradation assessment method is proposed. First, the self-organizing feature map (SOM) network is used to mine the latent degradation states of electrical machines. Second, the latent states are quantified according to established statistical health indexes, and by analyzing the distribution of extracted health indexes corresponding to different degradation states, the relevant transition relationships of the valid degradation states and the final evolving fault types are determined. Third, a hybrid discrete HMM is developed to fully describe the transition process among different states and assess the degradation scenario of a machine in an online manner. The results of a real application of an electric point machine show that the proposed method can identify valid degradation states and obtain a superior assessment accuracy.

## I. INTRODUCTION

ELECTRICAL equipment is widely used in various industries. Once a machine breaks down, it may reduce production efficiency and cause economic losses and accidents. From the initial deterioration to complete failure, a machine usually undergoes a series of degradation states. If the state of the equipment is promptly detected in this process, it can be inspected and maintained in a targeted manner to effectively prevent the occurrence of a failure. Condition-based maintenance (CBM) is a maintenance strategy that can optimize maintenance activities on the basis of condition monitoring (CM) data [1]. For a machine whose performance deteriorates continuously, CBM can provide predictive maintenance actions for its operating process and reduce the cost of degradation or failure by identifying and predicting its remaining useful life [2].

A prognostic-based maintenance technique [3] focuses on predicting the entire process of machine performance degradation rather than simply detecting, isolating, and determining the performance at a certain time. Compared with the fault-diagnosis technique, it can more effectively implement CBM [4]. In general, failure prognostics can be categorized into physical-model-based, data-driven, and hybrid approaches [5] – [7]. Physical model-based approaches combine a mathematical model based on a fault mechanism and measurement data to identify the model parameters and predict the degradation behavior. Data-driven approaches use information from collected data to identify the characteristics of the current degradation state and to predict future development trends. Hybrid approaches combine the first two methods to improve the prediction performance, although they are not yet mature [8]. Compared with data-driven approaches, physical-model-based approaches can use relatively less data and provide more accurate results. However, for a complex machine system, the establishment of precise and complete physical models is difficult. In these cases, data-driven approaches can serve as a solution. They only require a small amount of prior knowledge about the machine, e.g., the failure mechanism and expert experience, based on the CM data through various analysis techniques to mine the implicit information for assessment, which offers a tradeoff in terms of the complexity, cost, precision, and applicability [9].

Recently, data-driven performance degradation assessment has attracted increased attention, and some methods have been proposed. Data-driven approaches are generally divided into two categories: artificial intelligence (AI) and statistical approaches [5]. Neural network is one of the most commonly used AI approaches. Because of its powerful learning ability, it is used as a recognizer to distinguish the health state of a machine. A self-organizing feature map (SOM) is an unsupervised neural network, and its basic idea can be described as a nonlinear and smooth mapping of high-dimensional input space onto a low-dimensional array of neurons. Owing to its topology-preserving property, input data with similar features can be mapped to nearby neuron regions in the output space. An SOM is usually used for pattern recognition and clustering analysis. In terms of degradation assessment, Yu [10] used normal samples to train an SOM and calculated the health indication of CM data on the basis of this baseline model by comparing the difference between the indications to identify whether the machine is undergoing degradation. German et al. [11] trained an SOM using fault samples and identified the health state of a machine by analyzing the regions in the CM data mapped by a model. Hu et al. [12] used an SOM to cluster the CM data and determined the changing trajectory of the health states of a machine by analyzing the mapping results. As a statistical method, a hidden Markov model (HMM) is widely used for performance degradation assessment. An HMM is an extension of the Markov model. In this model, the stochastic process of a state transition is unobservable, and states can only be evaluated using a random process generated by another set of observations. This dual HMM stochastic process is similar to the degradation process of a machine to some extent. In practice, the health state of a machine is not observable and can only be indirectly observed through CM data. Therefore, an HMM can be used to describe the dual random process of health states and observations. Tonon-Mejia et al. [9] proposed a prognostic method based on a mixture of Gaussian HMMs (MoG-HMM) and used an MoG-HMM to model the machine-degradation behavior to evaluate and predict its health state. Jiang et al. [4] proposed a performance degradation assessment method in which a trained continuous HMM (CHMM) was used to identify abnormal features and evaluate the performance degradation of a machine. Yu [2] proposed an online learning framework based on an adaptive HMM (AHMM), which was used to learn the dynamic changes in the machine health, for degradation detection and adaptive assessment.

Through a literature review, we note that data-driven based performance degradation assessment approaches can generally be divided into three main steps: data feature processing, health index (HI) establishment, and performance evaluation. Feature processing consists of two parts: feature extraction and feature selection. Feature extraction converts CM data into a set of features that can describe the characteristics of a machine. Feature selection selects a subset of features with statistical significance from the feature set. HI establishment calculates the feature subset using appropriate algorithms and models to obtain indicators for measuring the machine degradation. Performance evaluation analyzes the HIs of the CM data, determines the current health state of a machine, and predicts the remaining useful life or future performance degradation trend over time. The methods reported in the literature, however, suffer from two obvious deficiencies in the performance evaluation: 1) determining the relevant relationships among diverse machine degradation states and assessing the specific degree of deterioration and 2) determining the evolutionary relationship among degradation and failure modes and assessing the failure modes corresponding to different degradation scenarios.

On the basis of an SOM neural network and improved HMM, a performance degradation assessment method is proposed in this paper to overcome the two deficiencies of previous work. In this method, feature processing on the measured power signal is first performed to obtain a low-dimensional feature set that describes the characteristics of an electrical machine. SOMs are used to cluster the feature set to obtain the latent degradation states. Multiple statistical HIs are established to quantify the latent states. The deterioration and evolutionary relationship among latent states are analyzed using the HI distribution to determine the valid degradation states. Hybrid discrete HMMs (Hybrid DHMM) are used to build state transition models for an online degradation assessment of the considered electrical machine. The merits of the proposed method are summarized as follows. 1) It can mine degradation states with different characteristics and determine the relevant relationship and degree of deterioration among these states. 2) It can determine the fault state of the final evolution of the degradation state. 3) It develops a new DHMM for the online assessment of electrical machine performance degradation. Compared with other improved DHMMs [13] – [15], this model can fully describe the transition process among normal-degradation-fault states and accurately identify the health state of a machine.

The remainder of this paper is organized as follows. Section II briefly introduces a theoretical background of the SOM and DHMM. Section III presents the proposed methodology for machine performance degradation assessment. Section IV examines a detailed case study using the power-signal datasets of an electric point machine. Concluding remarks are provided in Section V.

## II. THEORETICAL BACKGROUND

### A. SOM
#### 1) SOM Structure

SOM is a self-organizing feature mapping neural network proposed by Teuvo Kohonen in the 1980s [16], which is composed of a fully connected neuron array. It is characterized by its ability to cluster input data with different features by self-organizing and adaptively adjusting the network structure and parameters. Currently, SOM is widely used in the field of data mining [12], [17], [18]. This network consists of input and output layers. The number of input-layer neurons is $m$, which is the same as the dimension of the sample vector $X = \{x_i | i = 1, \cdots, m\}$, and the number of output-layer neurons is $n$, which is distributed in a two-dimensional (2D) array. These two neuron layers are fully connected by variable weight vectors $\omega_{ij}(t)$, $i = 1, \cdots, m$, $j = 1, \cdots, n$, where $\omega_{ij}$ is the weight between the input-layer neuron $i$ and the output-layer neuron $j$ and varies with time $t$. Therefore, an $m$-dimensional vector $w_j(t)$ of weights is associated with each neuron $j$ in the output layer. Through an unsupervised learning process, the output-layer neurons of an SOM are sensitive to the features of the different input vectors, and specific neurons act as recognizers of different input vectors. After network learning, all input samples are divided into different regions with respect to the neurons in the output layer; then, the clustering analysis of a dataset is completed.

*2) Clustering Algorithm*

The SOM learning algorithm is summarized as follows [19].

1. Initialization: When $t = 0$, small initial random values are assigned to the connection weights between two neuron layers. The initial learning rate of the network is set as $\eta(0)$. The initial adjacent neuron set of a output layer neuron $j$ is $N_j(0)$, and the number of iterations is $T$.

2. Input: The dataset sample vector $X(t)$ is sent to the input-layer neurons.

3. Competition. The Euclidean distance between the input vector $X(t)$ and the weight vectors of each neuron in the output layer is calculated as follows:

$$d_j = \sqrt{\left|\sum_{i=1}^{m}(x_i(t) - \omega_{ij}(t))^2\right|}, \ j = 1,\cdots,n. \quad (1)$$

The neuron with the smallest distance, called the winning neuron $j^*$, is obtained, which satisfies

$$d_{j^*} = \min_j(d_j). \quad (2)$$

4. Adaptation. The weight vectors of $j^*$ and its neighboring neurons are modified to make it approach $X(t)$:

$$w_j(t+1) = w_j(t) + \eta(t)(X(t) - w_j(t)), \ \forall j \in N_{j^*}, \quad (3)$$

where $\eta(t) \in (0,1)$ is the learning-rate function. As time $t$ varies, the rate gradually decreases until it approaches zero to ensure convergence of the learning process.

5. Judgment. Set $t = t + 1$. If $t < T$, steps 2–4 are repeated; otherwise, iteration is complete.

## B. DHMM

*1) HMM Components*

HMM is a probabilistic statistical model used to describe the transition and representation probabilities of hidden states of a system, which was proposed by Baum et al. in the 1960s [20]. HMMs have been widely used in the fields of pattern recognition [21] and fault diagnosis [13] – [15] owing to their good mathematical basic theory and improved algorithm. In most cases, it is difficult to detect the machine's health states directly; thus, they need to be inferred from observations. Therefore, the use of an HMM to describe a state transition is logical.

A complete HMM consists of five parts that can be represented by $\lambda = (N, M, \pi, A, B)$.

1. $N$: the number of states in the Markov chain. We assume that $S_1, \cdots, S_N$ are the $N$ states of the model, $q_t$ is the state in which the model remains at moment $t$, and $q_t \in (S_1, \cdots, S_N)$.

2. $M$: the number of observations per state. We assume that $V_1, \cdots, V_M$ are $M$ distinct observations, $O_t$ is the observation at moment $t$, and $O_t \in (V_1, \cdots, V_M)$.

3. $\pi$: the initial state probability vector. We assume that $\pi = (\pi_1, \cdots, \pi_N)$, where $\pi_i = P(q_t = S_i), \ 1 \le i \le N$.

4. $A$: the state transition probability matrix. We assume that $A = (a_{ij})_{N \times N}$, where $a_{ij} = P(q_{t+1} = S_j | q_t = S_i), \ 1 \le i,j \le N$.

5. $B$: the observation symbol probability matrix. We assume that $B = (b_{jk})_{N \times M}$, where $b_{jk} = P(O_t = V_k | q_t = S_j)$, $1 \le j \le N, \ 1 \le k \le M$.

The HMM state-transition process consists of two parts: one is the Markov chain that describes the transition of hidden states represented by $A$, and the other part is a stochastic process that describes the correspondence between the states and observations and is represented by $B$. For convenience, the notation $\lambda = (\pi, A, B)$ is used to represent the complete parametric setting of an HMM. According to the representation of the observation, HMMs can be categorized into CHMMs and DHMMs. CHMMs use a continuous probability density function to represent the observations of each state, whereas the observations of each state of a DHMM are represented by discrete vector sequences. When the two models describe the same state-transition process, the complexity and computational cost of DHMMs are relatively low [22]. Therefore, a DHMM is selected for degradation process modeling and health state recognition in the present work.

*2) Training Problems of DHMMs*

When a DHMM is built, because the parameters $(\pi, A, B)$ are unknown, the model needs to be trained first, and the health state of the machine is identified according to the obtained model parameters. Once the DHMM is used for modeling, the following three problems must be solved.

1. Detection: Given the observation sequence $O = o_1, \cdots, o_n$ and model $\lambda$, we compute the likelihood probability $P(O|\lambda)$ of the sequence given the model.

2. Decoding: Given the observations sequence $O$ and model $\lambda$, we find the hidden state sequence that is most likely to produce the observation sequence.

3. Learning: The model parameters $(\pi, A, B)$ are tuned to maximize $P(O|\lambda)$.

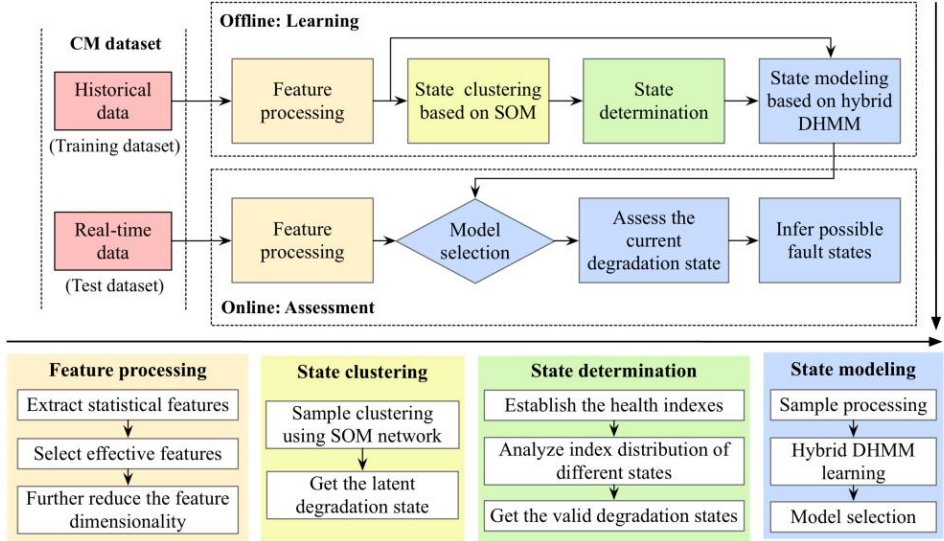

Fig. 1. System framework of the proposed methodology.

The three abovementioned problems can be effectively solved using the forward–backward [23], Viterbi [24], and Baum–Welch algorithms [25], respectively.

### III. DESCRIPTION OF THE PROPOSED METHOD

In this paper, a performance degradation assessment method based on SOM and Hybrid DHMM is proposed. This method relies on two main phases, as shown in Fig. 1: the offline learning and online assessment phases. The first phase consists of four parts. 1) Feature processing: the low-dimensional feature datasets are obtained through feature extraction, selection, and further dimensionality reduction of the monitored signal. 2) State clustering: a clustering analysis of the feature datasets based on the SOM is performed to obtain the latent-state datasets. 3) State determination: the HI distribution of the latent state data is used to determine the valid degradation states. 4) State modeling: the full-process state models of the electrical machine are built on the basis of the hybrid DHMM. The second phase assesses the performance degradation of a machine using the learned models. The processed real-time data are continuously fed to the learned models. By calculating the likelihood probability, a model that can best represent the observation data in the learned models is selected, and the corresponding state of the model is considered as the current health state of the machine. If this state is a degradation state, we can infer that the final state of its evolution represents the possible failures of the electrical machine in the future. In the proposed method, the SOM is applied to mine the degradation data of the electrical machine, and the hybrid DHMM is used to identify the degradation state. Obviously, the offline phase is the core in which the quality of the data and models directly affects the assessment results in the online phase.

### A. Feature Processing

A power signal can be obtained by calculating the voltage and current signals collected in real time. This type of machine signal contains a great deal of information about the electrical machine, which can describe the changes in the electrical parameters and forces during the running process. In addition, because the statistical features of the power signal change with the machine state in terms of the numerical value and distribution, the statistical feature data can provide clear information about the health state. In this study, the statistical feature data of the power signal are used to assess the performance degradation of electrical machines.

*1) Feature Extraction*

To fully reflect the health state of an electrical machine, we need to obtain as many statistical features of the power signal as possible. In this work, 10 statistical features (listed in Table I) are extracted from the time domain, which were used in [26] and [27]. $t_{i,m}$, $i=1,\cdots,P$, $m=1,\cdots,10$ is the $m$th statistical feature extracted from the $i$th time-domain phase of signal sample $x$. $P$ is the number of time-domain phases of the power signal. However, for some degradation and fault states, their symptomatic features are concentrated in specific value segments. If the signal features are only extracted from the time domain, part of the important features that describe the state of the machine can be omitted. To solve this problem, we project the signal curve to the value domain, analyze different power value domain in segments, and extract the eight statistical features listed in Table II from the value domain to fully describe the operating state of an electrical machine. $v_{j,n}$, $j=1,\cdots,S$, $n=1,\cdots,8$ is the $n$th statistical feature extracted from the $j$th value-domain segment of signal sample $x$, and $S$ is the number of value-domain segments of the power signal. By combining the time- and value-domain statistical features, the statistical feature vector $x$ of an electrical machine can be expressed as

$$x = \left(t_{1,1},\cdots,t_{1,10},\cdots,v_{S,1},\cdots,v_{S,8}\right). \tag{4}$$

The statistical feature dataset consisting of $N$ signal samples with dimension $D = 10P + 8S$ is $X = \{x_i | i = 1,\cdots,N\}$.

TABLE I
STATISTICAL FEATURES IN THE TIME DOMAIN

| Symbol | Feature name | Formula | Symbol | Feature name | Formula |
|---|---|---|---|---|---|
| $t_{i,1}$ | Out-to-in value | $P_i(end) - P_i(start)$ | $t_{i,6}$ | Sum of difference | $\sum_{c=1}^{C_i}(P_i(c+1) - P_i(c))$ |
| $t_{i,2}$ | Maximum difference | $\max P_i(c) - \min P_i(c)$ | $t_{i,7}$ | Kurtosis | $\dfrac{\frac{1}{C_i}\sum_{c=1}^{C_i}\left(P_i(c) - \frac{1}{C_i}\sum_{c=1}^{C_i} P_i(c)\right)^4}{(t_{i,5})^2}$ |
| $t_{i,3}$ | Mean value | $\frac{1}{C_i}\sum_{c=1}^{C_i} P_i(c)$ | $t_{i,8}$ | Crest factor | $\dfrac{\max P_i(c)}{\sqrt{\frac{1}{C_i}\sum_{c=1}^{C_i} P_i^2(c)}}$ |
| $t_{i,4}$ | Root mean square | $\sqrt{\frac{1}{C_i}\sum_{c=1}^{C_i} P_i^2(c)}$ | $t_{i,9}$ | Form factor | $\dfrac{\sqrt{\frac{1}{C_i}\sum_{c=1}^{C_i} P_i^2(c)}}{\frac{1}{C_i}\sum_{c=1}^{C_i} |P_i(c)|}$ |
| $t_{i,5}$ | Variance | $\frac{1}{C_i}\sum_{c=1}^{C_i}\left(P_i(c) - \frac{1}{C_i}\sum_{c=1}^{C_i} P_i(c)\right)^2$ | $t_{i,10}$ | Impulse factor | $\dfrac{\max P_i(c)}{\frac{1}{C_i}\sum_{c=1}^{C_i} |P_i(c)|}$ |

$C_i$ denotes the number of sample points in the *i*th phase, and $P_i(c)$ denotes the power value of a sample point $c$ in the *i*th phase.

TABLE II
STATISTICAL FEATURES IN THE VALUE DOMAIN

| Symbol | Feature name | Formula | Symbol | Feature name | Formula |
|---|---|---|---|---|---|
| $v_{j,1}$ | Maximum time value | $\max T_j(m)$ | $v_{j,5}$ | Median | $\text{media } P_j(m)$ |
| $v_{j,2}$ | Mean of segment value | $\frac{1}{M_j}\sum_{m=1}^{M_j} P_j(m)$ | $v_{j,6}$ | Maximum of segment value | $\max P_j(m)$ |
| $v_{j,3}$ | Data points | $M_j$ | $v_{j,7}$ | Time median | $\text{media } T_j(m)$ |
| $v_{j,4}$ | Maximum difference of segment | $\max P_j(m) - \min P_j(m)$ | $v_{j,8}$ | Mode | $\text{mode } P_j(m)$ |

$M_j$ denotes the number of sample points in the *j*th segment, $P_j(m)$ denotes the power value of a sample point $m$ in the *j*th segment, and $T_j(m)$ denotes the time value of a sample point $m$ in the *j*th segment.

### 2) Feature Selection

By extracting a large number of statistical features of the power signal, a high-dimensional feature dataset that reflects the health state of the electrical machine from various aspects is obtained. However, this feature set is not conducive to analysis for the following concerns: 1) the large number of data calculations, 2) the existence of features that cannot effectively represent the machine state, and 3) the redundancy among features. Because of these three problems, we need to select the features of the dataset and obtain an effective feature set that can characterize the state of the electrical machine. Over time, the degradation state of the machine will finally evolve into a fault state. During this process, the characteristic differences between the degradation and the fault states will gradually diminish and eventually disappear. Therefore, the fault features can be used as the basis for determining the degradation state of an electrical machine. Relative to the normal state, some features of the fault state will deviate from the normal value, and when the degree of deviation is greater, the discriminability of the features from the fault is more obvious. In this paper, the two-class feature selection method of the Fisher criterion [28] is used to analyze each of the fault and normal states to select effective features that can distinguish fault states.

We assume that $Fs = \{fs_i | fs_i \in \omega_j,\ i=1,\cdots,N,\ j=1,\cdots, C,\ \sum_j N_j = N\}$ is a $D$-dimensional typical fault feature dataset, where $N$ is the number of samples in the dataset, $\omega_j$ is the *i*th fault class, $C$ is the number of classes, and $N_j$ is the number of samples in class $\omega_j$. We further assume that $Ns = \{ns_i | i=1,\cdots,M\}$ is the $D$-dimensional normal state dataset, where $M$ is the number of samples in the dataset. $fs^{(d)}$, $y_i^{(d)}$, and $z^{(d)}$ respectively represent the values of the fault sample $fs$, the mean of the class $\omega_j$ fault-state samples, and the mean of the $Ns$ samples in the *d*th dimension. Then, the Fisher criterion value of the *d*th-dimensional feature of the class $\omega_j$ fault state is

$$F_i(d) = \frac{S_B^{(d)}}{S_W^{(d)}},\ d=1,\cdots,D, \tag{5}$$

where $S_B^{(d)}$ and $S_W^{(d)}$ respectively represent the inter- and intra-class variances in the *d*th-dimensional feature, which are calculated as follows:

$$S_B^{(d)} = \left(y_i^{(d)} - z^{(d)}\right)^2, \tag{6}$$

$$S_W^{(d)} = \frac{1}{N_j} \sum_{j \in \omega_i} \left( fs_j^{(d)} - y_i^{(d)} \right)^2 + \frac{1}{M} \sum_{j=1}^{M} \left( ns_j^{(d)} - z^{(d)} \right)^2 . \quad (7)$$

According to the Fisher criterion, if the criterion value is relatively large, the corresponding feature is more capable of distinguishing state classes. However, if two features within a class are linearly related and have relatively larger Fisher criterion values, then both will be selected, resulting in redundancy. To solve this problem, we use the correlation coefficient to measure the redundancy between the intra-class features:

$$\rho_{pq} = \frac{\sum_{j \in \omega_i} \left( fs_j^{(p)} - y_i^{(p)} \right)\left( fs_j^{(q)} - y_i^{(q)} \right)}{\sqrt{\sum_{j \in \omega_i} \left( fs_j^{(p)} - y_i^{(p)} \right)^2} \sqrt{\sum_{j \in \omega_i} \left( fs_j^{(q)} - y_i^{(q)} \right)^2}} , \quad (8)$$

where the range of $\rho$ is between -1 and 1. If $|\rho|$ is larger, the redundancy of the $p$th- and $q$th-dimensional features of the class $\omega_i$ fault state is larger. If $\rho$ is zero, these two features are independent.

The feature selection steps in this study are summarized as follows:

1. We use (5) to calculate the Fisher criterion values of the $D$-dimensional statistical features between each fault state and each normal state.

2. For the Fisher criterion values obtained for each fault state, we use half of the maximum Fisher criterion value as the standard value. We retain features whose criterion values are larger than the standard value and discard features whose criterion values are smaller than the standard value.

3. For the features retained for each fault state, we use (8) to calculate the correlation coefficient between each pair of two features. If the correlation coefficient is larger than 0.95, the feature with a smaller Fisher criterion value is removed.

*3) Feature Dimensionality Reduction*

By selecting the effective feature set, the non-effective and redundant features can be eliminated, and a concise feature set can be obtained. For a model, learning from a high-dimensional dataset can lead to an overfitting problem and affect the recognition performance [29]. Considering the effect of the feature-set dimensionality on the performance of the SOM and HMM, a further reduction in the dimensionality of the feature set that has been selected is necessary so that the models can be learned using a low-dimensional feature set to enhance their generalization and recognition performance. For the characteristics of the high dimension and nonlinearity of the power signal, a kernel principal component analysis (KPCA) [30], which is a nonlinear mapping method based on a kernel function, is used to further reduce the feature set in this study.

*B. State Clustering*

Using CM technology, the power-signal dataset of an electrical machine in different health states can be collected. By removing the fault samples from the dataset, a non-fault dataset containing both the normal and latent degradation state samples can be obtained. Categorizing the samples in this dataset, removing the normal state samples, and only retaining the latent state samples are key issues. In this work, the SOM is used for multiple clustering of the non-fault dataset. Through a comprehensive analysis of each clustering result, the samples in the non-fault dataset are selected, merged, and removed to obtain a new dataset that only contains the latent degradation state.

*1) Clustering Sequence*

A low-dimensional input dataset $X_{nf} = \{x_i \mid i = 1, \cdots, M\}$ can be obtained by performing feature processing on the non-fault dataset. The ensure the rationality of the clustering results, we use the SOM with different specifications to cluster this dataset to obtain the corresponding labels of the samples for different clustering results. By integrating the labels, the clustering sequence of each sample in the dataset is obtained. According to the clustering sequence, fast localization of the category of the sample can be realized. The steps to obtain the sample-clustering sequence are described as follows.

1. We set the SOM output-layer neurons to be arranged in a 2D array of $sn \times sn$. The number of neurons in the input layer is the same as the feature dimensionality of the sample in $X_{nf}$, and the first clustering of $X_{nf}$ is performed after the setting is completed. Then, we count the label number of the output-layer neurons corresponding to the sample in $X_{nf}$ and obtain the clustering category set $Clusa = \{a_{1,i} \mid i = 1, \cdots, M\}$, where $a_{1,i}$ is the neuron label number corresponding to the $i$th sample of $X_{nf}$.

2. We set the SOM output-layer neurons to be arranged in a 2D array of $(sn + 1) \times (sn + 1)$ and perform the second clustering of $X_{nf}$. Then, we count and obtain the clustering category set $Clusb = \{b_{1,i} \mid i = 1, \cdots, M\}$, where $b_{1,i}$ is the neuron label number corresponding to the $i$th sample of $X_{nf}$.

3. We set the SOM output-layer neurons to be arranged in a 2D array of $(sn + 2) \times (sn + 2)$ and perform the third clustering of $X_{nf}$. Then, we count and obtain the clustering category set $Clusc = \{c_{1,i} \mid i = 1, \cdots, M\}$, where $c_{1,i}$ is the neuron label number corresponding to the $i$th sample of $X_{nf}$.

4. We integrate $Clusa$, $Clusb$, and $Clusc$ to obtain the clustering sequence set $Seq_1 = \{(a_{1,i}, b_{1,i}, c_{1,i})^T \mid i = 1, \cdots, M\}$ of $X_{nf}$, where $(a_{1,i}, b_{1,i}, c_{1,i})^T$ is the clustering sequence corresponding to the $i$th sample of $X_{nf}$.

*2) Latent-State Analysis*

Combined with the clustering sequence, a latent-state analysis strategy is proposed in this paper. By analyzing the neuron distribution of the SOM output layer and selecting, merging, and removing the samples in the non-fault dataset, a new dataset that contains only the latent degradation states is obtained. The steps for this strategy are described as follows.

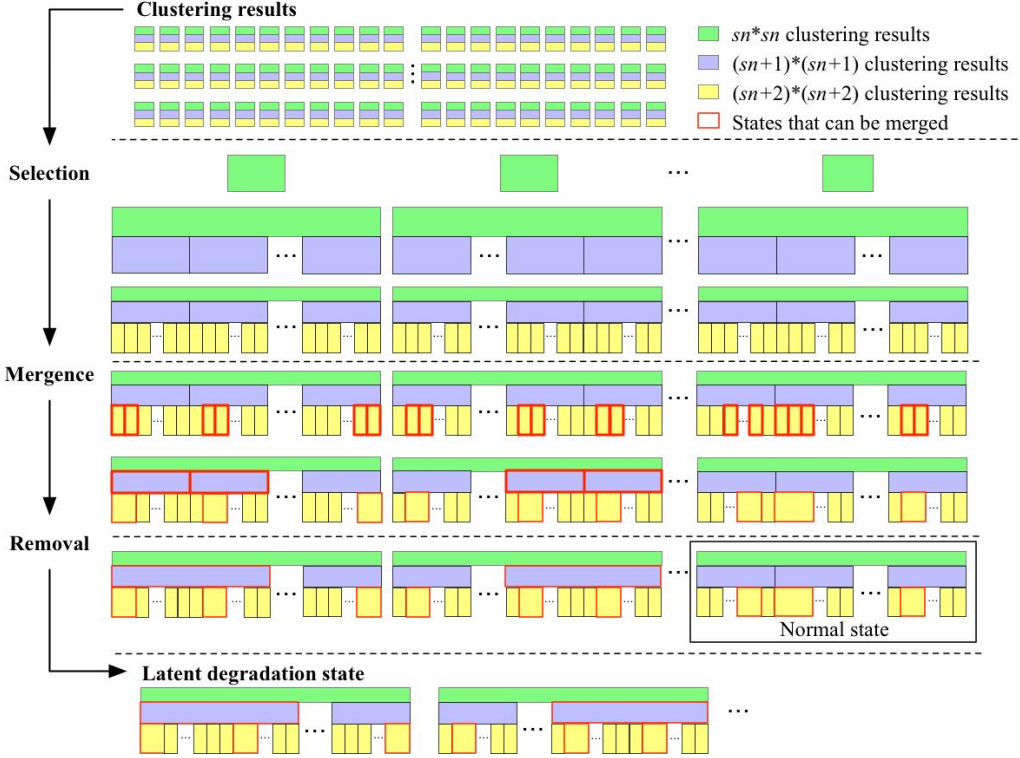

Fig. 2. Trajectory of the latent-state analytical strategy.

1. We count the label numbers of the $sn \times sn$ output-layer neurons whose number of clustering samples is not less than $M / sn \times sn$ and obtain the label number set $nNum_1$. According to the first-dimensional data of $Seq_1$, we select the samples corresponding to the label number of $nNum_1$ in $X_{nf}$. Then, the dataset $Sel_1 = \{x_i' | i = 1, \cdots, P\}$ and the corresponding clustering sequence set $Seq_2 = \{(a_{2,i}, b_{2,i}, c_{2,i})^T | i = 1, \cdots, P\}$ are constructed, where $P$ is the number of samples.

2. We count the label numbers of the $(sn+1) \times (sn+1)$ output-layer neurons whose number of clustering samples is not less than $M / (sn+1)*(sn+1)$ and obtain the label number set $nNum_2$. According to the second-dimensional data of $Seq_2$, the samples corresponding to the label number of $nNum_2$ in $Sel_1$ are selected. Then, the dataset $Sel_2 = \{x_i'' | i = 1, \cdots, Q\}$ and the corresponding clustering sequence set $Seq_3 = \{(a_{3,i}, b_{3,i}, c_{3,i})^T | i = 1, \cdots, Q\}$ are constructed, where $Q$ is the number of samples.

3. We count the label numbers of the $(sn+2) \times (sn+2)$ output-layer neurons whose number of clustering samples is not less than $M / (sn+2)*(sn+2)$ and obtain the label number set $nNum_3$. According to the third-dimensional data of $Seq_3$, the samples corresponding to the label number of $nNum_3$ in $Sel_2$ are selected. Then, we construct the dataset $Sel_3 = \{x_i''' | i = 1, \cdots, R\}$ and the corresponding clustering sequence set $Seq_4 = \{(a_{4,i}, b_{4,i}, c_{4,i})^T | i = 1, \cdots, R\}$, where $R$ is the number of samples.

4. According to the neuron distance distribution of the $(sn+2) \times (sn+2)$ output layer, we analyze each neuron neighborhood in $nNum_3$. The neurons that are nearest to these neurons in the neighborhood are merge, and the label numbers are consistent with the neurons with the largest number of clustering samples. By referring to the merged neuron label numbers, the third-dimensional data of $Seq_4$ are modified to update the corresponding clustering sequences of the $Sel_3$ samples and obtain $Seq_5 = \{(a_{5,i}, b_{5,i}, c_{5,i})^T | i = 1, \cdots, R\}$.

5. According to the neuron distance distribution of the $(sn+1) \times (sn+1)$ output layer, we analyze each neuron neighborhood in $nNum_2$. The neurons that are nearest to these neurons in the neighborhood are merged, and the label numbers are consistent with the neurons with the largest number of clustering samples. By referring to the merged neuron label numbers, we modify the second-dimensional data of $Seq_5$ to update the corresponding clustering sequences of the $Sel_3$ samples and obtain $Seq_6 = \{(a_{6,i}, b_{6,i}, c_{6,i})^T | i = 1, \cdots, R\}$.

6. According to $Seq_6$, the samples in $Sel_3$ are collated to their respective clustering states. We analyze the power-signal data of each state and remove the samples in the normal state to obtain the candidate feature dataset $Deg = \{ds_i | ds_i \in \xi_i,$ $i = 1,\cdots,S,\ j = 1,\cdots,E,\ \sum_j S_j = S\}$, where $S$ is the number of samples in $Deg$, $E$ is the number of latent states, $\xi_i$ is the $i$th latent state, and $S_j$ is the number of samples in the $j$th latent state.

The latent-state analysis strategy proposed in this paper is shown in Fig. 2. By analyzing the neuron distribution of the SOM output layer, the clustering results are selected, merged, and removed to obtain the candidate feature dataset that contains only the latent degradation states.

### C. State Determination

A clustering analysis of the non-fault dataset based on the SOM is performed, and the candidate feature dataset is obtained. Although the dataset is composed of several latent degradation state samples, the latent state only differs from the normal state in some features and may not be a valid degradation state of the electrical machine. Therefore, we need to further analyze the latent state and determine whether it is a truly valid degradation state. In this study, multiple HIs are established to analyze the degree of deterioration between the latent states and their final evolving fault states to determine the validity of these states.

#### 1) Multi-HI Set

In this study, multiple features of the power signal are used as HIs to discriminate the latent degradation state. Therefore, the feature data that characterize the machine state should be comprehensively analyzed to obtain the optimal feature set to determine the overall degradation process of the machine. The steps to establish the multi-HI set are as described as follows.

1. Equation (6) is used to further calculate the Fisher criterion value of the statistical feature of each dimension between every two fault states in the dataset $Fs$.

2. The maximum Fisher criterion value between every two fault states is counted, and the statistical feature corresponding to this value is used as an HI.

3. The maximum Fisher criterion value between each fault state and each normal state obtained in Section III.B.1 is counted, and the statistical feature corresponding to this value is used as an HI.

4. All of the obtained HIs are collected and arranged in sequence according to the statistical features of the vector in (5) to obtain the set $hI = \{idx_i | i = 1,\cdots,T\}$, where $idx_i$ is the feature dimension corresponding to the $i$th HI, and $T$ is the number of acquired HIs.

#### 2) Valid-State Determination

By analyzing the features of the different dimensions between the fault and normal states, an optimal feature set is obtained, which is used to determine the validity of the latent degradation states in the candidate dataset. In this work, the machine states are quantified using all of the HIs in this set, and the corresponding index distribution is obtained. By comparing the index distributions of the latent, typical fault, and normal states, we confirm the degree of deterioration of the latent states and their final evolving faults and complete the determination of the valid degradation states. Before the machine states are analyzed, the dataset needs to be normalized so that the feature values of each dimension of the sample can be distributed between zero and one. The normalization formula is expressed as follows:

$$(z')^{(d)} = \frac{z^{(d)} - min^{(d)}}{max^{(d)} - min^{(d)}}, \qquad (9)$$

where $(z')^{(d)}$ is the value of the $d$th-dimensional attribute of sample $z$ after normalization, $z^{(d)}$ is the $d$th-dimensional feature value of sample $z$, and $min^{(d)}$ and $max^{(d)}$ are the minimum and maximum feature values of the dataset in the $d$th dimension, respectively. We normalize the candidate feature dataset $Deg$, typical fault dataset $Fs$, and normal dataset $Ns$ using (9) to obtain $Deg'$, $Fs'$, and $Ns'$, respectively. According to the index set $hI$, we select the feature dimension of $Deg'$ and calculate the HI attribute values of various latent-state samples in the dataset using

$$md_{j,k} = \frac{1}{S_j} \sum_{i \in \xi_j} (ds'_i)^{(idx_k)},\ j = 1,\cdots,E,\ k = 1,\cdots,T, \qquad (10)$$

where $md_{j,k}$ is the $k$th HI attribute value of the $j$th latent state and $ds'_i$ is the $i$th sample in $Deg'$. After the calculation, we obtain the index attribute value set $mDeg = \{md_{j,k} | j = 1,\cdots,E,\ k = 1,\cdots,T\}$ of the latent states. Similarly, the HI attribute values of each state samples in $Fs'$ are calculated as follows:

$$mf_{j,k} = \frac{1}{N_j} \sum_{i \in \omega_j} (fs'_i)^{(idx_k)},\ j = 1,\cdots,C,\ k = 1,\cdots,T, \qquad (11)$$

where $mf_{j,k}$ is the $k$th HI attribute value of the $j$th fault state and $fs'_i$ is the $i$th sample in $Fs'$. Thus, the index attribute value set $mFs = \{mf_{j,k} | j = 1,\cdots,C,\ k = 1,\cdots,T\}$ of the fault states can be obtained. The HI attribute values of the samples in $Ns'$ are calculated using

$$mn_k = \frac{1}{M} \sum_{j=1}^{M} (ns'_i)^{(idx_k)},\ k = 1,\cdots,T, \qquad (12)$$

where $mn_k$ is the $k$th HI attribute value of the normal state and $ns'_i$ is the $i$th sample in $Ns'$. Then, the index attribute value set $mNs = \{mn_k | k = 1,\cdots,T\}$ of the normal state can be obtained. Through the abovementioned processes, quantification of the latent degradation, failure, and normal states of an electrical machine is completed. Further, according to the HI distributions of the states in $mDeg$, $mFs$, and $mNs$, the validity of a latent state is determined by the following steps.

1. Compare the HI distributions of the latent and fault states to determine the fault type corresponding to the latent states. If the latent state has a corresponding fault state, it is considered a valid degradation state; otherwise, it is treated as an invalid state and removed.

2. Determine the degree of deterioration of the degradation

state for the same fault type. By comparing the HI distributions of the normal and degradation states, the relevant state group is established, ranking the states within the group according to the degree of deterioration.

3. Analyze the signal curves and distribution of the index attribute values of each state in all relevant state groups to verify the validity of the degradation states.

*D. State Transition Modeling*

By analyzing the HI distribution of the latent degradation states of an electrical machine, the valid degradation states and relevant state groups are obtained. In the relevant state group, although the degree of deterioration of each state is obviously different, a reversible state transition will occur because of the influence of the work environment, routine maintenance, and other factors. Therefore, we need to describe the transition processes among the degradation states and build a corresponding model. In this study, a hybrid DHMM is proposed to investigate the transition processes among the degradation states, and a normal-degradation-fault full-process state model is developed to identify the health state of an electrical machine.

*1) Hybrid DHMM*

The complexity of a Markov chain leads to different DHMM structures. At present, the common structures are mainly divided into two types: ergodic and left–right types. However, these two structural models cannot completely describe the transition processes between the normal-degradation-fault states of electrical machines. In this paper, a new hybrid DHMM is proposed for full-process state modeling. Fig. 3 shows the hybrid DHMM of three relevant degradation states, where "0" denotes a normal state, "1" denotes a mild degradation state, "2" denotes a moderate degradation state, "3" denotes a severe degradation state, "4" denotes a fault state, $a_{ij}$, $i, j = 0, \cdots, 4$ denotes the transition probability among states, and $O_i$, $i = 0, \cdots, 4$ denotes the state observation. Compared with the ergodic- and left–right-type-structured DHMMs, this hybrid model of parallel and cross structures can describe the reversible transition process of the normal-degradation state and two degradation states and represent the one-way transition process of the degradation-fault state. For the obtained relevant state groups, the hybrid DHMM can be used for modeling to fully reflect the changing process of a machine health state during operation.

For some sequence-related modeling problems, the state-transition probability matrix *A* can reflect the information of the process state sequence. In the abovementioned model, *A* is represented as

$$A = \begin{bmatrix} a_{00} & a_{01} & a_{02} & a_{03} & 0 \\ a_{10} & a_{11} & a_{12} & a_{13} & a_{14} \\ a_{20} & a_{21} & a_{22} & a_{23} & a_{24} \\ a_{30} & a_{31} & a_{32} & a_{33} & a_{34} \\ 0 & 0 & 0 & 0 & a_{44} \end{bmatrix} \quad (13)$$

and has the following relationships:

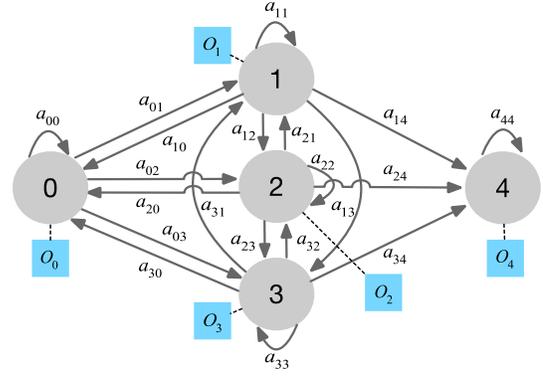

Fig. 3. Structure of the hybrid DHMM.

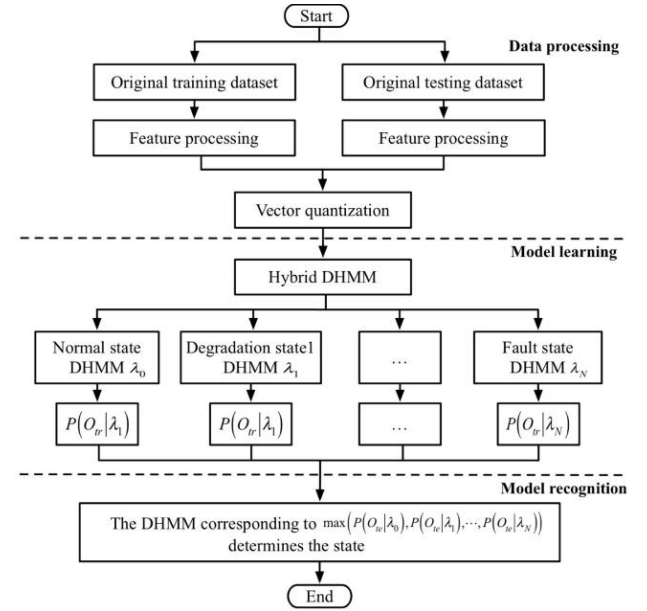

Fig. 4. Flowchart of the full-process state modeling based on a hybrid DHMM.

$$\begin{cases} 0 \leq a_{00} = 1 - a_{01} - a_{02} - a_{03} \leq 1 \\ 0 \leq a_{11} = 1 - a_{10} - a_{12} - a_{13} - a_{14} \leq 1 \\ 0 \leq a_{22} = 1 - a_{20} - a_{21} - a_{23} - a_{24} \leq 1 \\ 0 \leq a_{33} = 1 - a_{30} - a_{31} - a_{32} - a_{34} \leq 1 \\ a_{44} = 1 \\ 0 \leq a_{ij} \leq 1, \ i, j = 0, \cdots, 4 \end{cases} \quad (14)$$

Let us suppose that the probabilities of a machine in states $S_0$–$S_4$ are $p_0(t) - p_4(t)$, respectively, and that the probability distribution of the initial state is $\pi = [\pi_0(0), \pi_1(0), \pi_2(0), \pi_3(0), \pi_4(0)]$. Then,

$$P(t) = [p_0(t), p_1(t), p_2(t), p_3(t), p_4(t)] = \pi A^{(t)}, \quad (15)$$

where $A^{(t)}$ is a *t*-step state-transition probability matrix. According to the Chapman–Kolmogorov equation, we know that

$$A^{(t)} = (A^{(1)})^t = A^t. \quad (16)$$

Thus, (15) can be rewritten as

$$P(t) = [p_0(t), p_1(t), p_2(t), p_3(t), p_4(t)] = \pi A^t. \quad (17)$$

Because the initial state of the machine is generally a normal

TABLE III
DESCRIPTION OF THE ELECTRIC POINT MACHINE

| Name | Rated parameter | Name | Rated parameter |
| --- | --- | --- | --- |
| Power | 0.4 kW | Stroke | 150/220/240 mm |
| Voltage | Three-phase 380 V | Action time | $\leq 6.6$ s |
| Action current | $\leq 2$ A | Converter force | 3000–6000 N |
| Single-wire resistance | $\leq 54\ \Omega$ | Trailable force | 9/16/24/30 kN |

state, we have
$$\pi = [\pi_0(0), \pi_1(0), \pi_2(0), \pi_3(0), \pi_4(0)] = [1, 0, 0, 0, 0]. \quad (18)$$
The distribution of the states at any time can be obtained using (13), (17), and (18). It can be seen that this hybrid DHMM can fully reflect the state-changing process of the machine.

*2) Full-Process State Modeling*

The procedure for building the full-process state model based on the hybrid DHMM is shown in Fig. 4. It is mainly divided into the following three steps:

1. Data processing. The normal state samples, the degradation state samples in the relevant state group, and the evolving fault state samples are divided into training and test datasets according to an appropriate ratio. Feature processing on these two datasets is performed to obtain the corresponding low-dimensional datasets. Because the DHMM requires the input of discrete observation sequences, vector quantization is applied to the trained and test datasets in this study using the *k*-means algorithm [31], and the corresponding sequences $O_{tr}$ and $O_{te}$ are obtained.

2. Model learning. This learning step is essentially the process of solving the parameter-estimation problem of the model, and the objective is to obtain the optimal DHMM parameters $(\pi, A, B)$ for each state model of an electrical machine. The training sample sequence $O_{tr}$ of the *i*th ($i = 0, \cdots, N$) state is input, the initial state model $\lambda_{i0} = (\pi_0, A_0, B_0)$ is constructed, and the Baum–Welch algorithm is used to calculate a new model $\bar{\lambda}_i = (\bar{\pi}, \bar{A}, \bar{B})$ based on $\lambda_{i0}$. At this point, the likelihood probability $P(O_{tr} | \bar{\lambda}_i)$ of the $\bar{\lambda}_i$ output $O_{tr}$ is larger than $P(O_{tr} | \lambda_{i0})$ of the $\lambda_{i0}$ output $O_{tr}$. The Baum–Welch algorithm is used to iteratively calculate the parameters of the previous model until the optimal model $\lambda_i = (\pi, A, B)$ is obtained. The likelihood probability $P(O_{tr} | \lambda_i)$ of this model is the maximum and satisfies the convergence condition.

3. Model recognition. The forward–backward algorithm is used to calculate the likelihood probability of the output $O_{te}$ of each model $\lambda_i$ in the DHMM obtained from the learning step, and the model state corresponding to the maximum probability is used as the recognition result. In this work, test samples are used to verify the recognition performance of the hybrid DHMM to prove the feasibility of using it for online assessment.

## IV. EXPERIMENT AND ANALYSIS OF THE RESULTS

In this study, the power signal of a Siemens S700k AC point machine is used as the experimental data to verify the effectiveness of the proposed method. Point machines are important signal infrastructure used to change the position of a railway turnout. Table III summarizes the parameters of this machine. The real-time power calculation equation of the point machine is $P = U \cdot I \cdot \cos\theta$, where $\theta$ is the angle between the phase voltage $U$ and the phase current $I$. The workflow of the power-monitoring system is shown in Fig. 5. The switch amount acquisition module obtains the start and end times of the point machine by determining the state of the 1st turnout section relay (1DQJ). During this period, the Hall current sensor collects the current signals of the point machine at the 21st, 41st, and 61st nodes of the open-phase protector, and the output terminals of the Hall sensor are connected to the power collector. The power collector collects the voltage signals of each phase at the 11th, 31th, and 51th nodes of the open-phase protector. In addition, the sampling frequency of the current and voltage is 25 Hz. After the power signal is obtained by calculating the current and voltage signals, the power collector transmits the power data to the communication front-end processor via the bus and finally sends the power data to the monitoring host through the switch.

### A. Signal Sample Processing

By investigating the actual working conditions of the CM system at Chenzhou West Railway Station, we choose the 76-day power-signal samples of the point machine corresponding to the No. 1 turnout (J1, J2, J3, X1, and X2), No. 2 turnout (J1, J2, J3, X1, and X2), No. 3 turnout (J1), No. 4 turnout (J1), No. 7 turnout (J1, and J2), No. 11 turnout (J1, and X1), and No. 15 turnout (X1), which are monitored by the system, as the experimental data in this current work. According to the field maintenance log, we collate the signal data of the different state types of the point machines and construct the normal, non-fault, and typical-fault datasets. The numbers of samples are 100, 1800, and 180, respectively, where the typical-fault dataset consists of six common fault-state samples. Because the number of data points in the obtained power signal sample is not less than 165, the dimensionality of the datasets becomes too high. These original power signals need to be processed to obtain the feature datasets with an appropriate dimensionality for subsequent analysis.

*1) Experimental Setup for Sample Processing*

For the normal state *NS* and six typical fault states *FS1–FS6* of the point machine, 20 samples from the normal dataset and

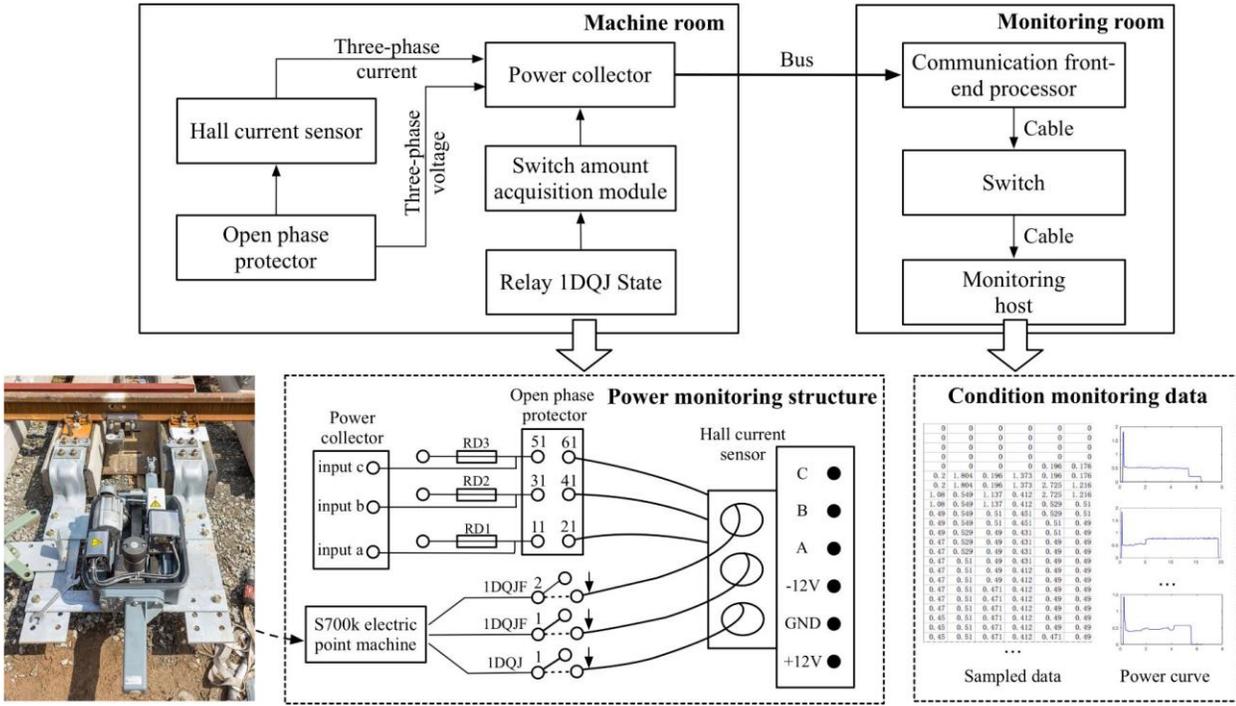

Fig. 5. Workflow of the point-machine monitoring system.

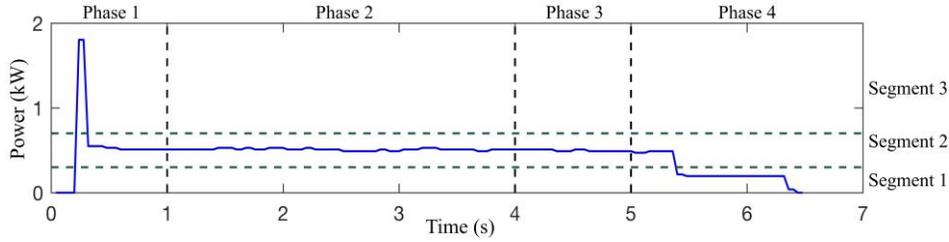

Fig. 6. Power range division of the power-signal curve.

TABLE IV
DESCRIPTION OF THE FAULT-STATE CHARACTERISTICS

| State | Characteristic |
|---|---|
| FS1 | In Phase 1, the power peak is extremely large, and the overall power value is larger than that in *NS*. |
| FS2 | In Phase 2, the degree of power fluctuation is large. |
| FS3 | In Phase 2, the power value sharply increases and remains constant until it reaches the time limit and then directly drops to zero. |
| FS4 | In Phase 3, the power value sharply increases and abnormally fluctuates until it reaches the time limit and then directly drops to zero. |
| FS5 | In Phase 4, the power value is approximately twice that in *NS*. |
| FS6 | In the later part of Phase 4, the power value drops and remains at zero. |

20 samples of each fault state from the fault dataset are selected in this work to create a standard dataset of 140 samples. Through the feature processing of this dataset, we obtain the basis for the division in the time-domain phase and value-domain segment, the selection of effective features, and the setting of the dimensionality-reduction parameters of the power signal. Subsequently, by referring to the abovementioned procedure, we perform feature processing on the non-fault dataset.

*2) Experimental Results for Sample Processing*

According to the working characteristics of the point machine, in this work, the time domain of the power signal is divided into four phases, as shown in Fig. 6: Phase 1 (0–1 s), Phase 2 (1–4 s), Phase 3 (4–5 s), and Phase 4 ($\geq 5$ s), and the value domain is divided into three segments: Segment 1 (0–0.3 kW), Segment 2 (0.3–0.7 kW), and Segment 3 ($\geq 0.7$ kW).

Fig. 7(a) shows the power curves of the normal and fault states, and Table IV lists the characteristics of the six fault states in the time domain. The power curve of the normal state and the six fault states are projected onto the value domain. Fig. 7(b) shows that the power value distributions of the machine fault states are significantly different. Compared with *NS*, the power peak of *FS1* is larger by approximately 5 kW, which is caused by the high action power in Phase 1. The power distribution of *FS2* is more discrete, which is caused by the large power fluctuation in Phase 2. *FS3* and *FS4* have more power points, which are caused by the long durations of Phases 2 and 3, respectively. In addition, most of the power values of these two faults are distributed between 0.5 and 1 kW, indicating that the power of the point machine in these two states is large. The power value of *FS5* has a specific distribution between 0 and 0.5 kW, which is caused by the abnormal change in power in Phase 4. *FS6* has a high power zero value, which is caused by the abnormal power value in

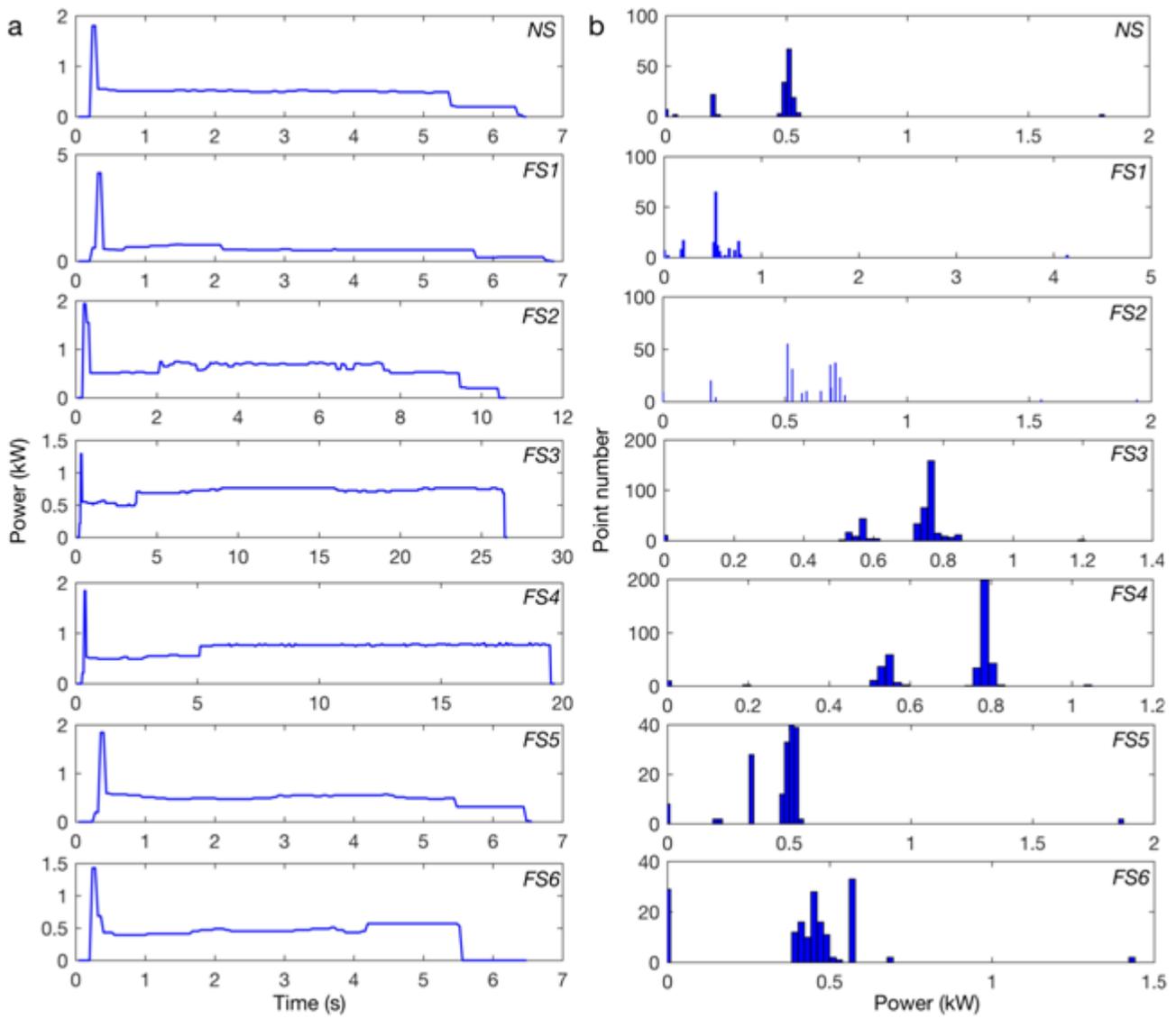

Fig. 7. Power signals of the normal and fault states. (a) Time-domain plots of *NS* and *FS1–FS6*. (b) Value-domain projections of *NS* and *FS1–FS6*.

Phase 4.

From Tables I and II, we extract the features of the standard dataset from the time and value domains and obtain a 64-dimensional feature dataset. Using this 64-dimensional dataset to select the effective features further forms a new 11-dimensional feature dataset from the features of $t_{1,3}$, $t_{2,1}$, $t_{2,2}$, $t_{2,8}$, $t_{4,3}$, $t_{4,5}$, $t_{4,8}$, $t_{4,9}$, $v_{1,2}$, $v_{1,7}$, and $v_{3,3}$, which correspond to the 4th-, 11th-, 12th-, 18th-, 33rd-, 35th-, 38th-, 39th-, 42nd-, 47th-, and 59th-dimensional features of the previous dataset. We normalize the 11-dimensional dataset and obtain the attribute value distribution of the samples in the selected dimensions, as shown in Fig. 8. We can see that the selected feature dimension can effectively distinguish the different states of the point machine, and no redundancy exists among these dimensions. Meanwhile, by retaining the effective features of the original data set, the dimension is reduced from 64 to 11 dimensions. We use KPCA to further reduce the dimensionality of the feature dataset, where we set the kernel function of the KPCA as a Gaussian radial kernel function and the number of principal components (PCs) is one to eight. By comparing the results of the different dimensionality reductions by the KPCA, we find that when the six PCs of the dataset are selected (which means that the dataset is reduced to six dimensions), the distinction among the different states is the best and can retain more than 99% of the original information. The attribute value distribution of the samples in the six-dimensional feature space is shown in Fig. 9.

From the abovementioned procedure, feature processing of the non-fault dataset is performed, and a six-dimensional feature dataset is obtained. Fig. 10 shows the attribute value distributions of the non-fault samples in each dimension. We can see that the distributions of the samples in the six dimensions are significantly different, and the characteristics of the samples can be distinguished.

### B. Degradation-State Validation

Through feature processing, a six-dimensional non-fault feature dataset is obtained. In this study, we use this dataset to mine the degradation states of the point machine. The mining includes two parts: clustering and determination. The clustering is based on the SOM to divide the non-fault samples into different state categories according to the similarity of the features. The determination uses HIs to perform a quantitative analysis of the state data and verify the validity of the state.

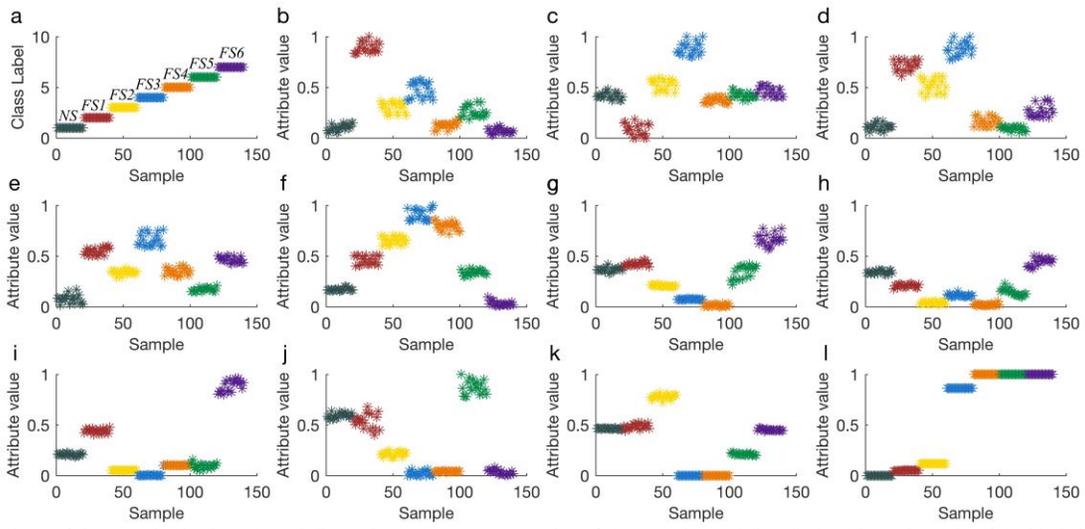

Fig. 8. Distributions of the samples in the selected dimensions: (a) state categories of the samples and (b)–(l) the 4th, 11th, 12th, 18th, 33rd, 35th, 38th, 39th, 42nd, 47th, and 59th selected dimensions, respectively.

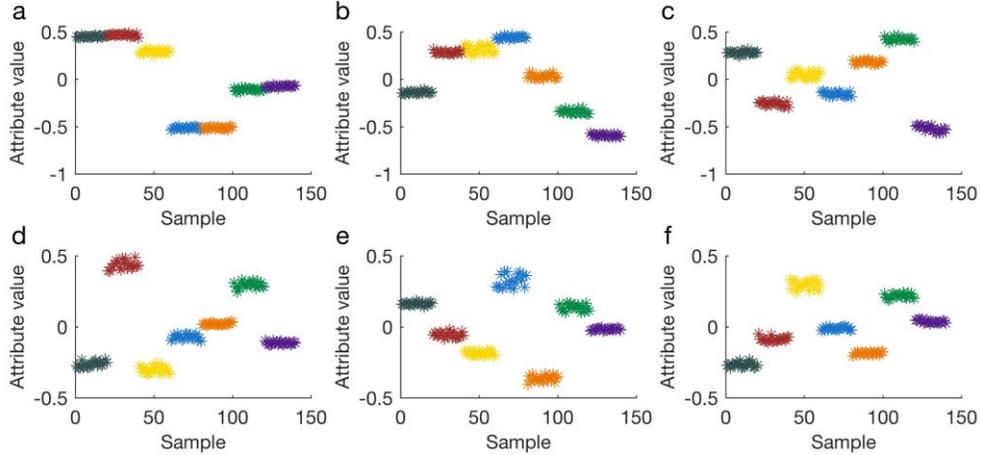

Fig. 9. Distributions of the samples in the six dimensions after KPCA transformation. (a)–(f) The first–sixth dimensions, respectively.

*1) Experimental Setup for Mining*

We use the SOM network to cluster the six-dimensional non-fault feature dataset, where the neurons are set in the network output layer arranged in 2D arrays of $4 \times 4$, $5 \times 5$, and $6 \times 6$. The neuron neighborhood shrinks in the shape of a hexagonal grid, the weights and biases of the network are randomly initialized, the number of iterations is 1000, the number of learning epochs is 20, and the loss function represents the mean squared error. After the completion of clustering, we analyze each neuron distribution in the SOM output layer and select, merge, and remove the dataset samples to obtain the candidate dataset.

We use the 64-dimensional fault feature dataset and 64-dimensional standard feature dataset to calculate the HIs. According to the HIs, we quantify the latent-state samples of the candidate dataset, the six state samples of the fault dataset, and the normal dataset samples to obtain the HI distribution in each state. We analyze these distributions to determine the validity of the latent states.

*2) Analysis of the Results*

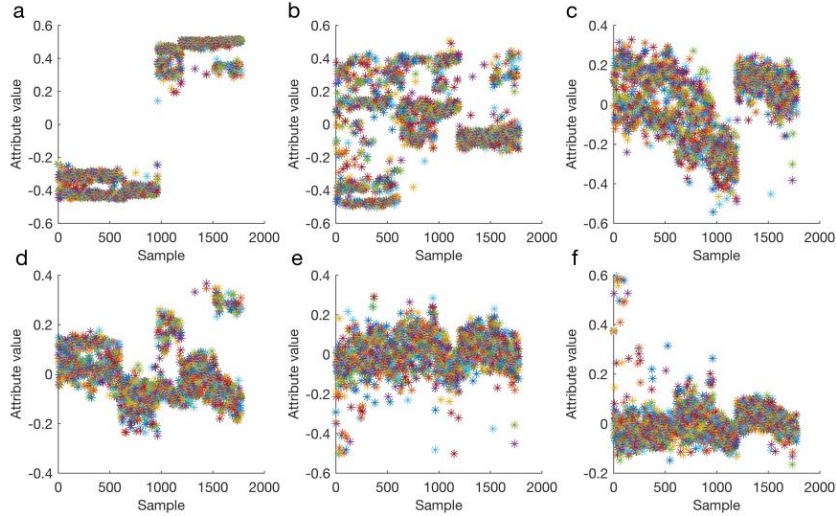

Fig. 10. Distributions of non-fault samples in six dimensions after feature processing. (a)–(f) The first–sixth dimensions, respectively.

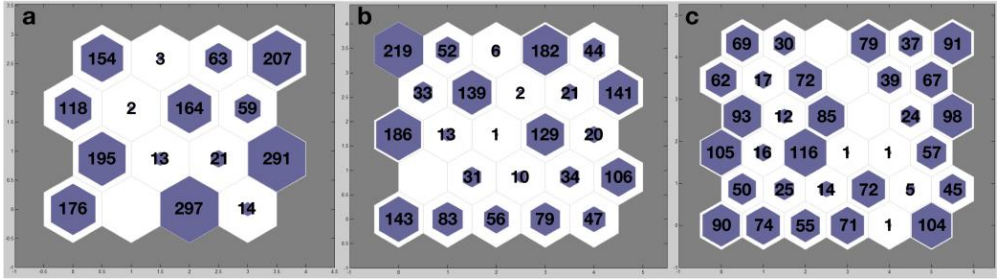

Fig. 11. Distributions of the samples in the SOM: (a) 4 ×4, (b) 5 ×5, and (c) 6 ×6 output layers.

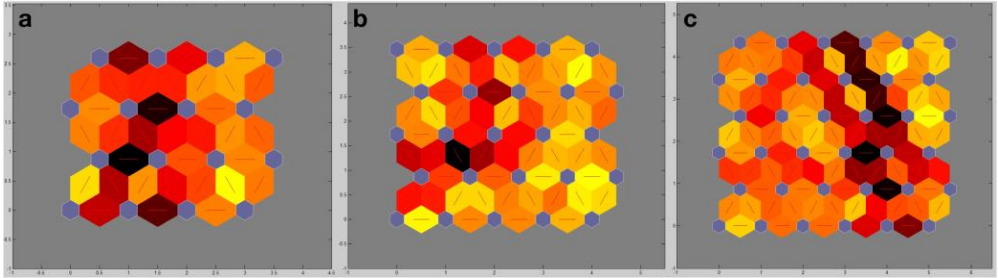

Fig. 12. Distributions of the neuron distance in the SOM: (a) 4 ×4, (b) 5 ×5, and (c) 6 ×6 output layers.

In this work, we use the Neural Network Toolbox™ developed by MathWorks, Inc. and select the SOM Neural Network of the toolbox to perform a clustering analysis of the six-dimensional non-fault feature dataset. The distributions of the samples after clustering are shown in Fig. 11, where the hexagonal lattice represents the neurons in the output layer, and the numbers on the lattice represent the number of samples clustered by the neurons. We can see that when the number of neurons in the output layer increases, the distribution of the samples among the neurons becomes more dispersed and cannot be concentrated under specific neurons. Therefore, the number of neurons in the output layer should be reasonably set to avoid the problem of poor clustering caused by a large number of neurons. The distribution of the neuron distance after clustering is shown in Fig. 12, where the connection band between adjacent neurons is used to measure the distance. A lighter color for the connection band represents a smaller distance, and a darker color represents a larger distance.

By selecting, merging, and removing samples from the non-fault dataset, six latent states, namely, *S1–S6*, are mined in this study. The statistical information of each state is listed in Table V, where the total number of latent samples is 556, which account for 31.2% of the dataset, and the number of normal samples is 201, which account for 11.3% of the dataset. These results indicate that more than half of the samples in the dataset are not divided into seven clustering states. During the transition from the normal state to the degradation state, the machine experiences several intermediate states, and the difference between the intermediate and normal states is indistinct. Compared with the degradation state, the features of the intermediate state are not typical for reflecting the health degradation. Thus, a large proportion of the intermediate-state samples in the dataset are not divided into clustering states.

*S2*, *S4*, and *S5* evolve into *FS2*, and the degrees of deterioration of these three latent states gradually increase, as shown in Fig. 14(c). 2) *S3* and *S6* evolve into *FS4*, and the degrees of deterioration of these two latent states gradually increase, as

TABLE V
STATISTICAL INFORMATION OF THE CLUSTERING STATES

| Clustering sequence | S1 | S2 | S3 | S4 | S5 | S6 | NS |
|---|---|---|---|---|---|---|---|
| SOM—4 ×4 | 16 | 5 | 9 | 11 | 8 | 13 | 3 |
| SOM—5 ×5 | 11 | 21 | 10 | 20 | 17 | 24 | 21 |
| SOM—6 ×6 | 1 | 10 | 24 | 19 | 13 | 35 | 15 |
| Sample size | 70 | 94 | 90 | 90 | 117 | 95 | 201 |

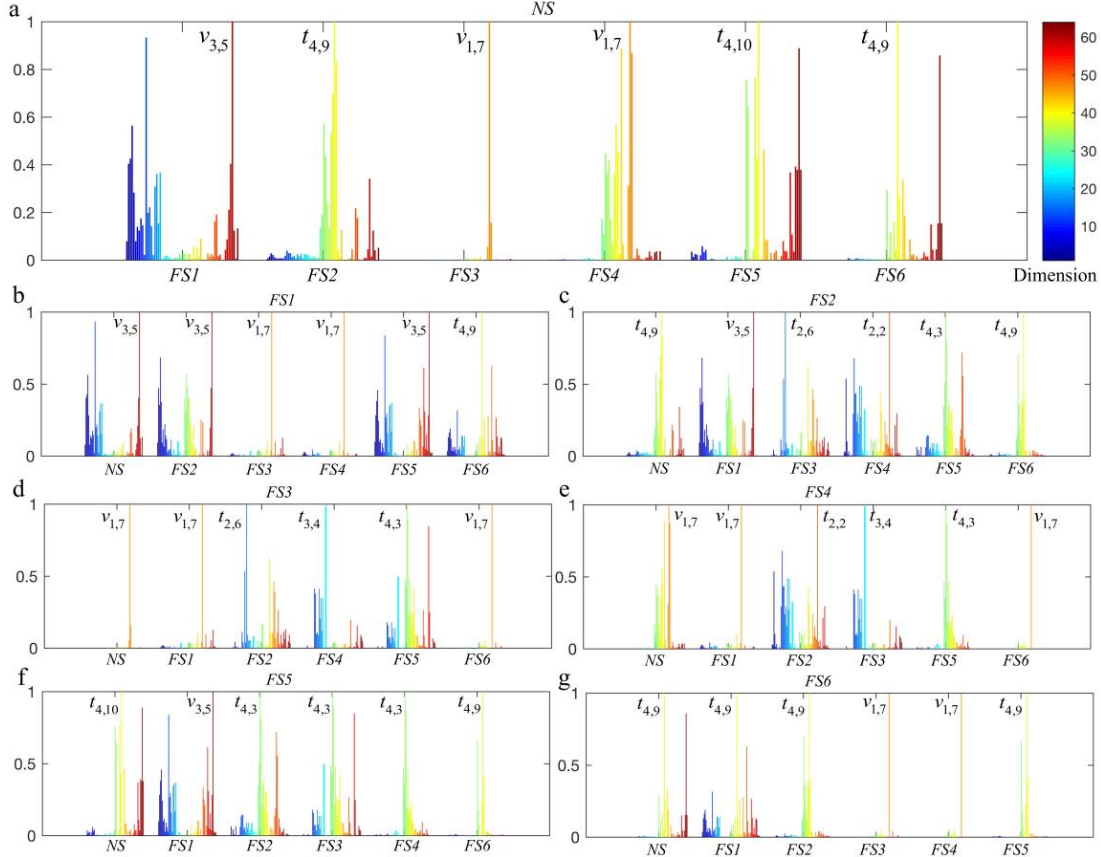

Fig. 13. Distributions of the Fisher criterion values among different states. (a) *NS* and other states. (b) *FS1* and other states. (c) *FS2* and other states. (d) *FS3* and other states. (e) *FS4* and other states. (f) *FS5* and other states. (g) *FS6* and other states.

Therefore, non-fault datasets with a rich-state type and large capacity should be used in the degradation mining by eliminating the intermediate-state samples of the dataset to obtain as many latent-state samples as possible.

We use the feature datasets to calculate the Fisher criterion values among the different states. The value distributions are shown in Fig. 13. By selecting the optimal feature as the HI, the multi-index set $\{t_{2,2}, t_{2,6}, t_{3,4}, t_{4,3}, t_{4,9}, t_{4,10}, v_{1,7}, v_{3,5}\}$ is obtained.

According to the multi-HI set, we calculate the HI attribute values of the normal state, the six latent states, and the six fault states and obtain the HI distribution for each state, as shown in Figs. 14(a) and (b). We can therefore determine the following. 1) shown in Fig. 14(d). 3) *S1* does not have a corresponding fault state and is considered invalid. Therefore, we select the two relevant state groups of the point machine, namely, *G1* = (*S2, S4, S5*) and *G2* = (*S3, S6*).

We verify the validity of *G1* and *G2*. 1) The sample curves of the three states in *G1* are shown in Fig. 15(a). In Phases 1, 3, and 4, the power value distributions of the three states are similar to that of *NS* with no obvious anomaly. In Phase 2, as shown in Fig. 15(b), the power values and the degrees of fluctuation of the three increase in turn, which are consistent with the failure characteristics of *FS2*. Further, through an analysis of Figs. 17(a) and (b), we can see that in the dimensions of the HI, namely, $t_{2,2}$ and $t_{2,6}$, the attribute values of the three state samples show stepwise growth characteristics, and the degree of deterioration gradually increases. 2) The sample curves of the two states in *G2* are shown in Fig. 16(a). In Phases 2 and 4, the power value distribution of the two state samples is normal. In Phase 1, both peak values are lower than

In addition, we use KPCA to directly reduce the normal-, degradation-, and fault-state feature datasets and extract the first three PCs of the samples to obtain a three-dimensional (3D) spatial distribution map, as shown in Fig. 18. Fig. 18 shows that that of *NS*, and the peak value of *S6* is higher than that of *S3*. In Phase 3, as shown in Fig. 16(b), the power values and the degrees of fluctuation of the two increase in turn, which are consistent with the failure characteristics of *FS4*. Further, by analyzing Figs. 17(a), (c), and (d), was can see that the attribute values of the two state samples also show stepwise growth characteristics in the dimensions of the HI, namely, $t_{2,2}$, $t_{3,4}$, and $t_{4,3}$.

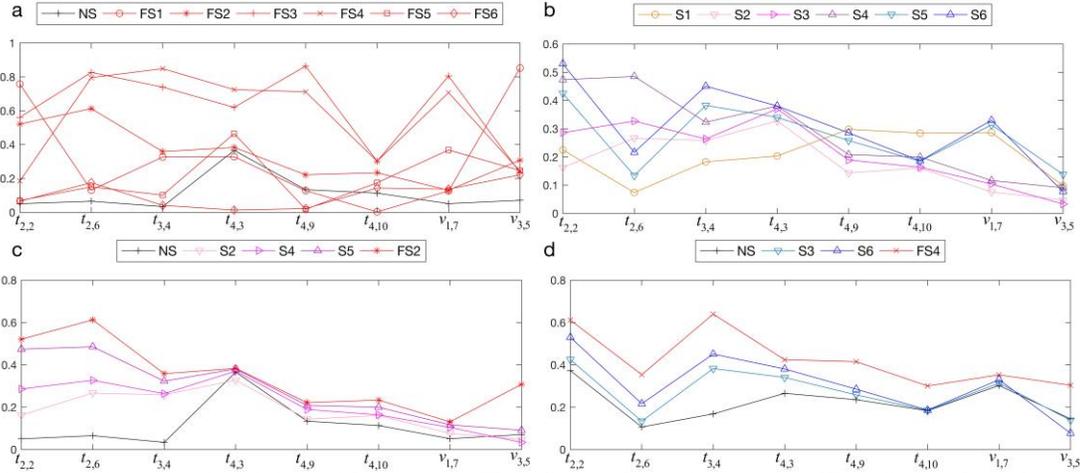

Fig. 14. HI distributions of the normal, degradation, and fault states. (a) HI distribution of *NS* and *FS1–FS6*. (b) HI distribution of *S1–S6*. (c) HI distribution of *NS, S2, S4, S5*, and *FS2*. (d) HI distribution of *NS, S3, S6*, and *FS4*.

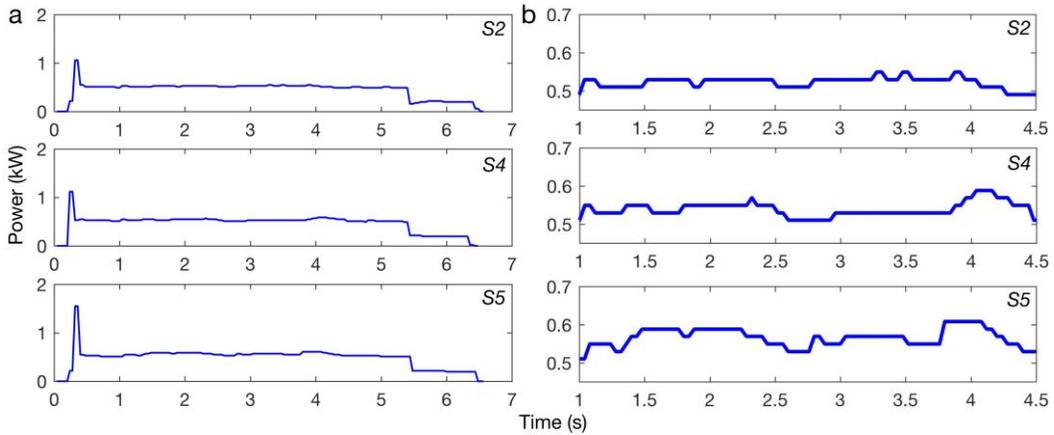

Fig. 15. Power curves of *G1*. (a) Global curves of *S2*, *S4*, and *S5*. (b) Local magnifications of *S2*, *S4*, and *S5* in Phase 2.

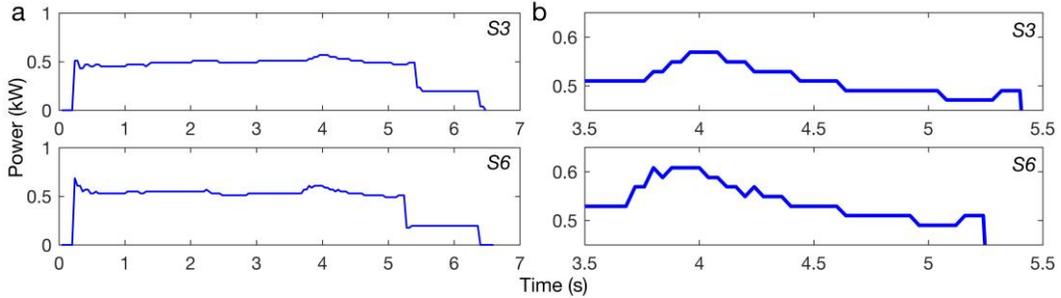

Fig. 16. Power curves of *G2*. (a) Global curves of *S3* and *S6*. (b) Local magnifications of *S3* and *S6* in Phase 3.

*G1* and *G2* are distributed between the normal- and evolving-fault-state samples, and the degree of deterioration of each state sample in the group increases, which further proves the validity of *G1* and *G2*.

### C. State Recognition Modeling

By mining the non-fault dataset, two relevant state groups of the point machine are obtained. For these two groups, we build the full-process state models based on the hybrid DHMM and verify their performance, which proves the feasibility of the models to identify the health state of the point machine.

observation sequence, where the $k$ value of the algorithm is 10.

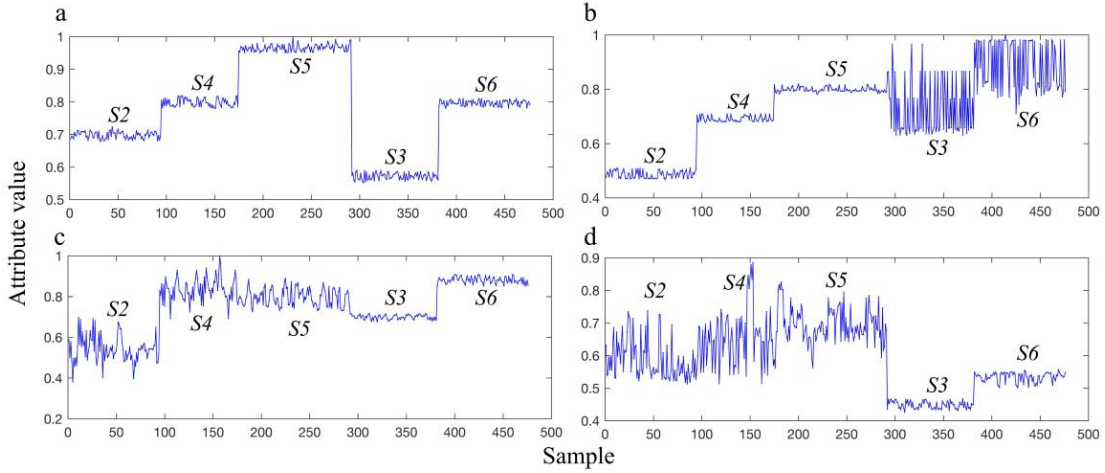

Fig. 17. Distributions of the degraded samples in the HI dimensions. (a) Sample attribute values in $t_{2,2}$. (b) Sample attribute values in $t_{2,6}$. (c) Sample attribute values in $t_{3,4}$. (d) Sample attribute values in $t_{4,3}$.

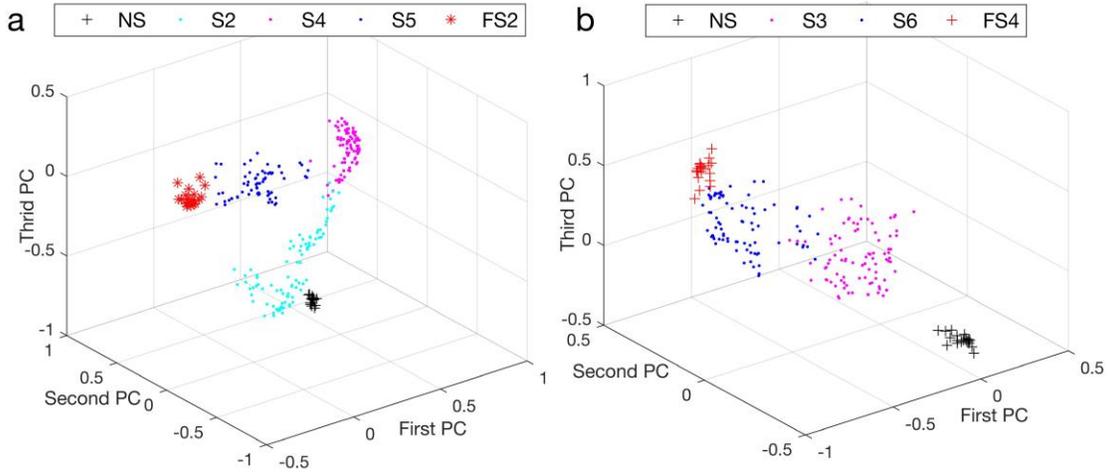

Fig. 18. Distributions of the samples in 3D space. (a) *NS*, *G1*, and *FS2* samples. (b) *NS*, *G2*, and *FS4* samples.

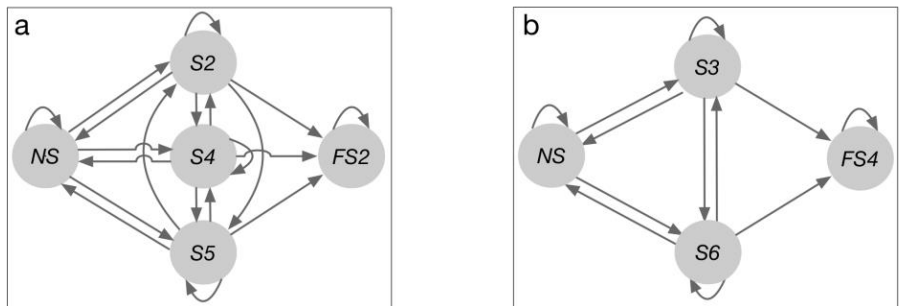

Fig. 19. Hybrid DHMM state transition relationships. (a) Hybrid DHMM$_1$. (b) Hybrid DHMM$_2$.

### 1) Experimental Setup for Modeling

We construct the dataset for *NS*, *G1*, and *FS2*. The number of samples in each state is 90, and the dataset contains a total of 450 samples. We randomly divide the samples in each state in the dataset according to the ratio 3:1 to construct the training and test datasets. We perform feature processing of these two datasets to obtain six-dimensional feature datasets. We quantify the feature datasets using the *k*-means algorithm to obtain the

We set the parameters of Hybrid DHMM$_1$, where the number of hidden states is six and the number of observations is 10. The transition relationship among the hidden states is shown in Fig. 19(a). The initial state probability vector is $\pi_1 = [1,0,0,0,0]$. We randomly generate the initial observation symbol probability matrix. Similarly, we construct the datasets for *NS*, *G2*, and *FS4*. The number of samples in each state is 90, and the dataset contains 360 samples. We set the parameters of Hybrid DHMM$_2$, in which the number of states is four. The transition relationship among the hidden states is shown in Fig. 19(b). The initial state probability vector is $\pi_2 = [1,0,0,0]$. Subsequently, we train and test Hybrid DHMM$_1$ and Hybrid DHMM$_2$ and analyze the recognition performance of the 15 iterations. The one-time recognition results of the hybrid model are shown in Fig. 20(b), and the recognition results for the different methods are listed in Table VI.

Figs. 21(a) and (b) respectively show the one-time training

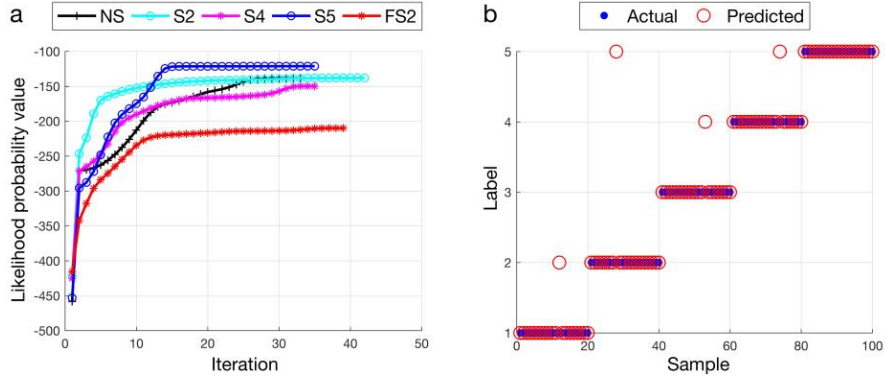

Fig. 20. Hybrid DHMM$_1$ training and validation. (a) Training curves. (b) Recognition results of the different states.

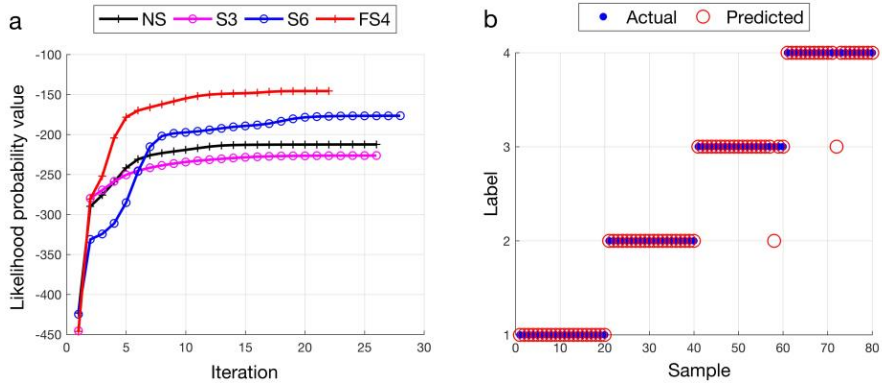

Fig. 21. Hybrid DHMM$_2$ training and validation. (a) Training curves. (b) Recognition results of the different states.

TABLE VI
COMPARISON OF THE RECOGNITION ACCURACIES OF THE FIVE STATES

| State categories | KPCA–Hybrid DHMM | PCA–Hybrid DHMM | LLE–Hybrid DHMM | KPCA–DHMM |
|---|---|---|---|---|
| *NS* (%) | 98 | 97 | 96 | 98 |
| *S2* (%) | 96 | 96 | 97 | 93 |
| *S4* (%) | 97 | 95 | 95 | 94 |
| *S5* (%) | 97 | 96 | 98 | 95 |
| *FS2* (%) | 100 | 98 | 99 | 96 |
| RA (%) | 97.6 | 96.4 | 97 | 95.2 |

TABLE VII
COMPARISON OF THE RECOGNITION RATES OF THE FOUR STATES

| State categories | KPCA–Hybrid DHMM | PCA–Hybrid DHMM | LLE–Hybrid DHMM | KPCA–DHMM |
|---|---|---|---|---|
| *NS* (%) | 100 | 98.5 | 98.5 | 96 |
| *S3* (%) | 97 | 96 | 95 | 95 |
| *S6* (%) | 97 | 95.5 | 97 | 94 |
| *FS4* (%) | 96 | 96 | 98 | 94 |
| RA (%) | 97.5 | 96.5 | 97.125 | 94.75 |

models.

*2) Analysis of the Model Performance*

Because of the randomness of the initial parameters, we train and verify each state model of the hybrid DHMM 10 times and take the mean of each state recognition result as the recognition accuracy (RA) of the hybrid model. The one-time training curves of the five state models of Hybrid DHMM$_1$ are shown in Fig. 20(a), which shows that all states basically converge after curves of the four state models and one-time recognition results of hybrid DHMM$_2$. Table VII lists the recognition results for the different experiment methods.

The lists in Tables VI and VII indicate that the RA of the hybrid DHMM based on the KPCA dimensionality reduction is the highest. Compared with PCA, KPCA can extract and retain the nonlinear information in the power signal. By using the kernel method to extend the PCA from linear to nonlinear, the KPCA implicitly considers the nonlinear structure of the signal features and preserves more original data information than the PCA. Locally linear embedding (LLE), as a nonlinear manifold learning method, can preserve the original topological structure of the signal features after dimensionality reduction [32]. Although the RA based on the LLE dimensionality reduction is slightly lower than that of the KPCA, it can also retain the nonlinear information of the original data. In addition, compared with the left–right–type DHMM, the hybrid DHMM can more accurately identify the health state, which proves the effectiveness of this improved HMM for the full-process state recognition of point machines.

## V. Conclusion

In this paper, a novel performance degradation assessment method was proposed for electrical machines. The three major contributions of this study are as follows: 1) degradation states with different characteristics are mined, and the relationships between these states are determined; 2) the fault states of the final evolution of the degradation states are determined; and 3) a new hybrid DHMM for the online assessment of electrical-machine performance degradation is developed. The effectiveness of the proposed method is verified using the CM datasets of an electric point machine, which contain a massive number of samples involving different health states under various operating conditions. By using the assessment results of these datasets, we show that the proposed method can mine multiple latent degradation states of machinery. By analyzing the HI distributions of different health states, we determine the relevant deterioration relationships between the degradation states and the final fault states. The full-process state model can effectively assess the performance degradation of machines.